\documentclass[useAMS,usenatbib]{mn2e} \usepackage{graphicx}
\usepackage{txfonts}
\newcommand{\msun}{{\rm M}_{\sun}}
\newcommand{\xte}{\textit{RXTE}}

\title[Soft-state power spectra of Cyg X-1] {Damped harmonic oscillator interpretation of the soft-state power spectra of Cyg X-1}  

\author[R. Misra and A. A. Zdziarski]{Ranjeev Misra$^1$\thanks{E-mail:
rmisra@iucaa.ernet.in (RM), aaz@camk.edu.pl (AAZ)} and Andrzej A. Zdziarski$^2$\footnotemark[1]\\
$^1$Inter University Center for Astronomy and Astrophysics, Pune University Campus, Pune 411007, India\\
$^2$Centrum Astronomiczne im.\ M. Kopernika, Bartycka 18, 00-716 Warszawa, Poland\\
}
\begin{document}

\date{Accepted 2008 April 5. Received 2008 March 30; in original form 2008 March 3}

\pagerange{\pageref{firstpage}--\pageref{lastpage}} \pubyear{2008}

\maketitle

\label{firstpage}

\begin{abstract}
We develop a model of an accretion disc in which the variability induced at a given radius is governed by a damped harmonic oscillator at the corresponding epicyclic frequency. That variability induces both linear and non-linear responses in the locally emitted radiation. The total observed variability of a source is the sum of these contributions over the disc radius weighted by the energy dissipation rate at each radius. It is shown that this simple model, which effectively has only three parameters including the normalization, can explain the range of the power spectra observed from Cyg X-1 in the soft state. Although a degeneracy between the black hole mass and the strength of the damping does not allow a unique determination of the mass, we can still constrain it to $\la (16$--$20)\msun$. We also show that our model preserves the observed linear rms-flux relationship even in the presence of the non-linear flux response.
\end{abstract}

\section{Introduction}

Black-hole binaries are primarily observed to be in two main spectral states (see, e.g., \citealt{Zdz04} for a review on the spectral states and the related variability). In the hard/low state, the spectrum can be approximately described as a hard power-law, with the typical photon index of $\Gamma \approx 1.7$, and a high energy cutoff above $\sim 100$ keV \citep{gierlinski97}, and only a weak soft X-ray component. The likely physical origin of that spectrum is thermal Comptonization of some soft photons by a hot plasma with $kT \sim 10^2$ keV \citep[e.g.,][]{gierlinski97,Fro01}. The hot plasma may be identified as an inner hot accretion disc surrounded by a truncated outer cold disc, which provides seed photons for Comptonization \citep{Sha76,Esi97,zdziarski02}. On the other hand, in the soft/high state,  the spectrum consists of a soft blackbody-like component, which dominates the luminosity, and a hard X-ray power-law tail with a photon index $\Gamma \approx 2.5$--3 with no detectable high energy cutoff. While the soft component is consistent with the emission of a cool accretion disc extending close to the innermost stable orbit, the nature of the hard power-law tail is less clear. Detailed modeling of spectra and long-term variability shows that it is well represented by Comptonization of soft photons by a hybrid plasma, i.e., containing both thermal and non-thermal electrons \citep{Pou98,Gie99,Zdz01,zdziarski02}. This plasma may exist in the form of active, time-varying, regions above the cold disc \citep{Haa93,Pou99}. In addition, during transitions between those two states, black-hole binaries also show the state called intermediate or very high. In each state, there are significant spectral and flux variations on a variety of time-scales.

The temporal variability of a black hole system is different in the hard and soft spectral states. During the hard state, the power spectrum, $P$, is approximately a broken power law with the breaks at the frequencies of $f \sim 0.1$ Hz and $f \sim 5$ Hz \citep{Now99,Gil99}, whereas the power spectrum in the soft state is approximately a power law with a high frequency cutoff above $f\sim 10$ Hz \citep*{Chu01}. In addition, discrete features, referred to as quasi-periodic oscillations (QPOs), are often observed. Attempts have been made to identify the centroid frequencies of these QPOs with frequencies at specific radii of the disc, which could either be the truncation radius \citep{gs04} or the radius of a resonance between the radial and vertical epicyclic frequencies \citep{abramowicz03}. These intriguing interpretations do not generally address the origin of the continuum variability. Detailed but empirical fitting of the power spectra using Lorentzian functions reveal that the continuum variability is complex in nature. Often three or more components (with a total of nine or more parameters) are required to adequately describe the spectra \citep{Now00}. However, it is possible that this apparent complexity appears because the Lorentzian function may not be an adequate description of the variability. In the framework of the active flare model, the continuum variability can be explained by the rise and fall time scales of the flares and the frequency at which they occur \citep{Pou99}. However, this interpretation does not make a direct connection with the characteristic dynamic frequencies of the accretion disc.

The accretion process in a disc occurs via angular momentum loss through turbulence, which nature and origin remain uncertain. In the standard accretion disc model \citep{Sha73}, this uncertainty is parametrized by approximating the turbulent behavior as an effective viscosity providing a stress proportional to the local pressure. While such a prescription may indeed provide a reasonable approximation to the time-averaged disc, the effect of turbulence on time-dependent properties of the disc is much more uncertain. \cite{Lyu97} showed that if the viscosity parameter varies at each disc radius with a local characteristic frequency, the accretion rate in the inner disc fluctuates with the flicker-noise behaviour, i.e., with $P \propto 1/f$. This provides a natural explanation for the observed power-law of the soft-state of black hole systems. Also, it explains the X-ray variability observed mostly at low frequencies in spite of the characteristic frequencies of the inner disc, where the X-rays originate, being much higher. Such a generic model where variability propagates from larger radii to smaller ones can also explain the observed frequency-dependent photon-energy lags in Cyg X-1 \citep{Mis00,Kot01}. It is also consistent with the observed linear rms-flux relationship \citep{uh01}. 

In these models, a high frequency break in the power spectrum is likely to correspond to a characteristic frequency at the innermost radius. Since the break observed in the soft state is typically at $\sim 10$ Hz, \cite{Lyu97} suggested that this corresponds to the viscous time scale of the inner disc.  \cite{Tit07} have generalized this model by invoking a power-law variation of the viscosity parameter with radius, and used a specific form for the local perturbations. They show that the observed power spectra of black hole systems can be explained  by a model with two components, corresponding to the outer disc and an inner hot region.

A crucial element of such models, as noted by \cite{Lyu97}, is that the time scale for the variations should be as long as the viscous one in order to explain the observed high frequency break. On the other hand, the main 
characteristic time scale is the epicyclic one, which is significantly shorter. \cite{Kin04} note that even the characteristic magnetic dynamo time scale is also expected to be significantly smaller than the viscous one. If the dissipation is driven by turbulence induced by magnetic instabilities, it is that dynamo time scale which should dominate the observed variability. \cite{Kin04} and \cite{May06} provide a solution to this problem by suggesting that large scale outflows can only occur when weak random events in different disc annuli by chance give rise to a coherent poloidal field. The time scale for that is much longer than the local dynamo one. It is then these large scale outflows that govern the observed slow variability. Those authors have developed a specific disc model incorporating the stochastic dynamo action as well as jet/wind outflows. Their numerical results suggest that it may also explain observed variability of black hole systems.

In all these models, the variability is associated with the unknown turbulence and not with the characteristic dynamic time scales of the disc. On the other hand, the {\it simplest\/} variability model would associate it with epicyclic oscillation at each radius, with the net variability due to a superposition of these oscillations. In this approach, even if variability at large radii induces some low frequency fluctuations in the inner regions \citep{Lyu97}, the dominant local power should correspond to the epicyclic time scales. However, 
such a natural model in its simple version would predict a much higher break frequency than that observed. 

In this work, we propose that these natural oscillations do indeed occur, but a damping, viscous-like, effect, suppresses the high frequencies. A likely reason for the viscous damping is the same turbulence that causes the viscous dissipation. Thus, contrary to the models mentioned above where the observed variability is primarily due to variations in the turbulent viscosity and/or dynamo effects, the variability in our model is due to the natural
dynamic oscillations of a disc but damped by viscous-like effects (perhaps also the turbulence). Based on this idea, we have developed a simple framework that explains the power spectra of black-hole binaries in the soft state. We limit the analysis to this state, where, as mentioned earlier, an optically-thick disc extends nearly to the last stable orbit. Although most of the observed variability in the soft state occurs in the high energy photons, and hence it should be associated with active regions or flares, it is expected that the underlying disc is still the main driver of the variability (e.g., via magnetic buoyancy). 

We assume that the disc variability at each radius is governed by a damped harmonic oscillator at the corresponding epicyclic frequency. This dynamic variation induces both linear and non-linear responses in the emitted radiation. The total observed variability of a source is then a sum of these contributions, weighted by the relative contribution of the radiation from a given radius. We find that the predicted power spectrum is completely determined by only three effective parameters including normalization, and it can explain the different spectral shapes observed for Cyg X-1 in the soft state. Whereas a degeneracy between the black hole mass and the damping factor does not allow the determination of the mass of the system, the model constrains it to $\la (16$--$20)\msun$, which is consistent with other mass measurements.

In Section \ref{model}, we describe the model. In Section \ref{fits}, we compare the predicted power spectra with observations from {\it Rossi X-ray Timing Explorer\/} (\xte). In Section \ref{rms-flux}, we discuss implications of our results for the relationship between the rms and the flux. In Section \ref{discussion}, we briefly discuss and summarize the results.

\section{The Model}
\label{model}

The variability of a dynamic quantity, $X(t)$, at each radii of the accretion disc is assumed to be governed by a damped harmonic equation,
\begin{equation}
\ddot  x(t) + f_{\rm E}^2(r) x(t) + f_{\rm D} (r)\dot x(t) = C(t),
\label{harmequ}
\end{equation}
where $x(t) = [X(t)-X_0]/X_0$ is the normalized fractional variation, with $X_0$ being an average of $X(t)$. Then $C(t)$ is a stochastic driving term and $f_{\rm E}(r)$ is the epicyclic frequency, which for the Schwarzschild metric equals \citep[e.g.,][]{Ste99},
\begin{equation}
f_{\rm E}(r) = \frac{1}{2\upi}\frac{c}{r_{\rm g}} \left(\frac{r}{r_{\rm g}}\right)^{-1.5} \left(1 -
\frac{6 r_{\rm g}}{r}\right)^{1/2},
\label{epicyclic}
\end{equation}
where $r_{\rm g} \equiv GM/c^2$ and $M$ is the mass of the black hole. Then $f_{\rm D}$ is an unknown damping frequency, which we parameterize as $f_{\rm D} (r) =\lambda f_{\rm E} (r)$ ($\lambda=0$ corresponds to no damping).  While we do not specify the nature of the dynamic normalized variable, $x(t)$, we assume that it induces a linear response departing from the average photon flux, $S_0$, such that $s(t) \equiv [S(t)-S_0]/S_0 \propto x(t)$.  Thus, the observed linear component of the power spectrum corresponding to an annular radius, $r$, is \citep{Lan69},
\begin{equation}
p_{\rm l}(f) = | \tilde s(f) |^2 \propto | \tilde x(f) |^2 \propto
\frac{|\tilde C(f)|^2}{(f_{\rm E}^2-f^2)^2+\lambda^2 f_{\rm E}^2 f^2},
\end{equation}
where $\tilde s(f)$, $\tilde x(f)$ and $\tilde C(f)$ are the Fourier transforms of $s(t)$, $x(t)$ and $C(t)$, respectively. It is assumed that in the frequency range of interest, the driving fluctuations, $C(t)$, are dominated by a $1/f$ noise such that $|\tilde C(f)|^2 \propto  f^{-1}$ valid for $f$ greater than a low frequency cutoff, $f_{\rm min}$. Thus,
\begin{equation}
p_{\rm l}(f) = \frac {\alpha^2}{I(f_{\rm E},\lambda)}\frac{f^{-1}}{(f_{\rm E}^2-f^2)^2+\lambda^2 f_{\rm E}^2 f^2},
\end{equation}
where the normalization,
\begin{equation}
I(f_{\rm E},\lambda) \equiv \int^\infty_{f_{\rm min}} \frac{f^{-1} {\rm d}
f}{(f_{\rm E}^2-f^2)^2+\lambda^2f_{\rm E}^2f^2},
\end{equation}
makes the fractional rms of the observed variability equal to $\alpha$. If $f_{\rm min}\ll f$ (which we assume hereafter), $I$ becomes,
\begin{equation}
I(f_{\rm E},\lambda) \simeq f_{\rm E}^{-4}\left[  \ln(f_{\rm E}/f_{\rm min}) + B(\lambda)\right],
\end{equation}
where 
\begin{equation}
B(\lambda) = \cases{[b(\lambda)/4][\pi +2\arctan b(\lambda)], &$\lambda < 2$;\cr
- 1/2,  &$\lambda = 2$;\cr
[b(\lambda)/4]\ln \frac{b(\lambda)-1}{b(\lambda)+1},    &$\lambda > 2$
,\cr} 
\end{equation}
and 
\begin{equation}
b(\lambda) = \frac{2-\lambda^2}{\lambda \sqrt{|4-\lambda^2|}}.
\end{equation}
Thus, the normalization of $p_{\rm l}(f)$ depends only logarithmically on $f_{\rm min}$, while the spectral shape is independent of it.

The total observed power spectrum, $P_{\rm l}(f)$, is then a weighted sum of the contributions from different radii,
\begin{equation}
P_{\rm l}(f) = \frac{\int^\infty_{r_{\rm in}} p_{\rm l}(f) G(r) 2 \upi r {\rm d}r} {\int^\infty_{r_{\rm in}} G(r) 2 \upi r {\rm d}r},
\label{weighted}
\end{equation}
where $G(r)$ is a weight function describing the relative contribution of each radius to the total variability, $p_{\rm l}$ is also dependent on $r$ via the epicyclic frequency, equation (\ref{epicyclic}), and on $\lambda$. We assume hereafter $r_{\rm in}=6 r_{\rm g}$. In writing equation (\ref{weighted}), it has been implicitly assumed that the driving fluctuations at different radii, $C(t)$, are incoherent. This assumptions requires that $r_{\rm coh} \ll r$, where $r_{\rm coh}$ is the radial coherence length of the underlying variability, which is justified if, e.g., $r_{\rm coh}$ is of the order of the scale height of the disc. Most of the variability observed in black-hole systems in the soft state is associated with the high energy component \citep{Chu01}, which probably arises from hot active regions on top of the cold disc. The radial extent of these active regions is not clear, and, in particular, the energy dissipated in them as a function of radius remains unknown. Thus, it is difficult to obtain the form of the weight function, $G(r)$,  from first principles. Here we make a simple assumption (also used for modeling disc rms spectra in \citealt{zdziarski05}) that $G(r)$ is proportional to the local energy dissipated in the cold disc, i.e.,
\begin{equation}
G(r) \propto r^{-3}\left[1-\left(\frac{6 r_{\rm g}}{r}\right)^{1/2}\right].
\label{G_r}
\end{equation}

\begin{figure}
\begin{center}
\includegraphics[width=0.9\linewidth,angle=0]{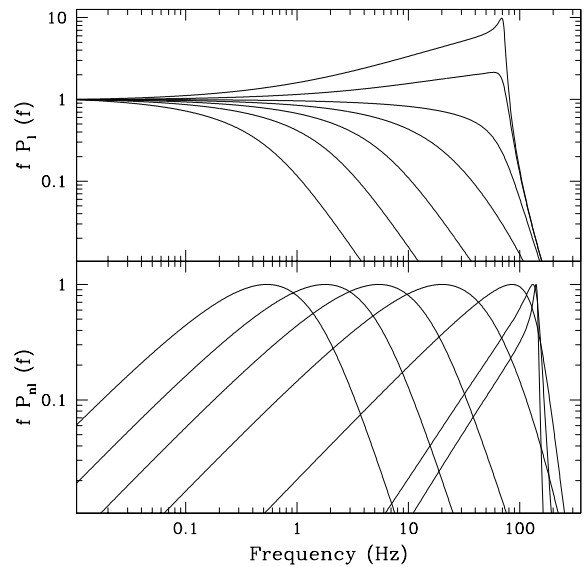}
\end{center}
\caption{The power per logarithm of frequency, $f P(f)$, for $M=10\msun$. The curves from right to left are for $\lambda = 0.1$, 0.3, 1, 3, 10, 30, 100, for (a) the linear power, $P_{\rm l}$, and (b), the non-linear one, $P_{\rm nl}$. The normalization of each curve corresponds to unity at (a) $f P_{\rm l}(0.01\,{\rm Hz})$, (b) the peak of $f P_{\rm nl}$.
}
\label{f:models}
\end{figure}

The observed spectrum, $P_{\rm l}(f)$, depends on the normalization factor, $\alpha$, the damping factor, $\lambda$, and, through $f_{\rm E}$, on the black-hole mass, $M$. Of these, $\alpha$ is the normalization, while variations in $M$ shift the spectrum in frequency. Hence, $\lambda$ is the
single parameter that describes the spectral shape in this linear model. In Fig.\ \ref{f:models}(a), we show $P_{\rm l}$ for several different values of $\lambda$ and $M=10\msun$, $f_{\rm min}=10^{-4}$ Hz. At small values of $\lambda\ll 1$, the oscillator is almost undamped, and the spectrum has a narrow spike at $\simeq 71(10\msun/M)$ Hz. This corresponds to the maximum epicyclic frequency, $f_{\rm E,m}$, equation (\ref{epicyclic}), which is achieved at $r \simeq 8 r_{\rm g}$, very close to the maximum of the gravitational energy dissipation rate, $G(r)$, equation (\ref{G_r}). In the time domain, we can define the corresponding time scale as $(2\upi f_{\rm E,m})^{-1}\simeq 2(M/10\msun)$ ms. Observationally, it roughly corresponds to the shortest e-folding time scales seen yet (which happened in the soft state) from Cyg X-1 \citep{gz03,Zdz04}. 

For $f \ll f_{\rm E,m}$, $P_{\rm l} \propto 1/f$, which is a consequence of the form of driving variability of $\tilde C(f) \propto 1/f$. As $\lambda$ increases, the system becomes more damped, and the spectrum shows a cutoff at a frequency $\ll f_{\rm E,m}$. For large values of $\lambda \ga 1$, an increase in $\lambda$ is nearly  equivalent to a scale shift of $P_{\rm l}(f)$ in frequency.  Such a frequency shift also occurs when the mass of the black hole is increased. This results in a degeneracy between the parameters, $M$ and $\lambda$, when the model is fitted to observational data (see Section \ref{fits}). Therefore, we can effectively determine only $\lambda/M$.
 
The response of the observed flux to the variability, $x(t)$, may
also have a significant non-linear component, which in the leading order
would be $s_{\rm nl} (t) \propto x^2(t)-\langle x^2(t)\rangle$. The Fourier transform of
$s_{\rm nl}$ can be expressed as,
\begin{equation}
\tilde s_{\rm nl} (f) \propto \int_{-\infty}^{\infty} \tilde x(f')
\tilde x (f-f') {\rm d} f'.
\end{equation}
Expressing $\tilde x(f) = \sqrt {p_{\rm l}} {\rm e}^{{\rm i}\phi(f)}$ and changing the variable to $\Delta f = f/2 - f'$, the non-linear term can be
rewritten as,
\begin{equation}
\tilde S_{\rm nl}(f)  \propto\! \int\limits_{-\infty}^{\infty} \!\!\! \sqrt{p_{\rm l} (f/2 -\Delta f)  p_{\rm l} (f/2 +\Delta f)} \hbox{e}^{{\rm i}\phi(f/2+\Delta f)-{\rm i}\phi(f/2-\Delta f)} {\rm d} \Delta f.
\end{equation}
Since the driving term $C(t)$ is stochastic, the phase, $\phi(f)$, will be largely incoherent. Thus, the above integral will average to zero except when $2\Delta f < \Delta f_{\rm coh}$, where $\Delta f_{\rm coh}$ is the frequency range over which the phase of $\tilde C(f)$ remains coherent. For small $\Delta f_{\rm coh} \ll f$, $\tilde S_{\rm nl} \propto p_{\rm l}(f/2) \Delta f_{\rm coh}$, and the non-linear power spectrum can be expressed as 
\begin{equation}
p_{\rm nl}(f) = \frac{\beta^2}{I_{\rm nl} (f_{\rm E},\lambda)} f^2 p_{\rm l}^2(f/2),\quad I_{\rm nl}(f_{\rm E},\lambda) = \int^\infty_{f_{\rm min}} \!\!\!f^2 p_{\rm l}^2(f/2) {\rm d}f,
\end{equation}
where $\Delta f_{\rm coh} \propto f$ has been assumed, $I_{\rm nl}$ is the normalization, and $\beta$ is the fractional rms.

The observed non-linear component of the  power spectrum, $P_{\rm nl}$, is then the weighted average of $p_{\rm nl}$ over the energy dissipation distribution, $G(r)$, analogously to equation (\ref{weighted}). Apart for the normalization factor, $\beta$, $P_{\rm nl}$ depends on $\lambda$ and $M$. Similar to the linear case, variation in $M$ results in an overall frequency shift, and the shape of the spectrum is determined only by $\lambda$. In Fig.\ \ref{f:models}(b), the computed spectra, $P_{\rm nl}$, are shown for different values of $\lambda$ and $M=10\msun$. Similar to the linear component, variation of $\lambda$ when $\lambda > 1$ results only in a frequency shift, which is degenerate with variation of $M$.

The total observed spectrum will, in general, be a combination of the linear, $P_{\rm l}$, and non-linear, $P_{\rm nl}$, components, with the total fractional rms of $\sqrt{(\alpha^2+\beta^2)}$.  Apart from this normalization factor, the spectrum depends on $\lambda$, the black hole mass, $M$, and the power ratio between the non-linear and linear components, $q_{\rm nl} = \beta/\alpha$.

\section{Fits to data}
\label{fits}

The Proportional Counter Array (PCA) on board \xte\/ has repeatedly observed the Cyg X-1. During some of the observations the system was in the soft state, and a criterion to select soft state data can be obtained by defining a soft colour, $C_{\rm s}$, as the ratio of the energy fluxes in the energy bands 4.0--6.4 keV to 3.0--4.0 keV \citep{dg03}. The system exhibited a range of $0.7 \la C_{\rm s} \la 2.4$, and in this work we define the system to be in the soft state when $C_{\rm s} \la 1.2$. We have selected 11 pointed observations (during 1996--2004) satisfying this criterion and covering the range of $0.7\la C_{\rm s} \la 1.2$ approximately uniformly (M. Gierli\'nski, personal communication). Table \ref{t:fits} gives the IDs of the selected observations. Rebinned power spectra were calculated from the light curves using standard techniques. A 5 per cent systematic error was added in quadrature to the statistical error of the spectra. 

\begin{figure}
\begin{center}
{\includegraphics[width=1\linewidth,angle=0]{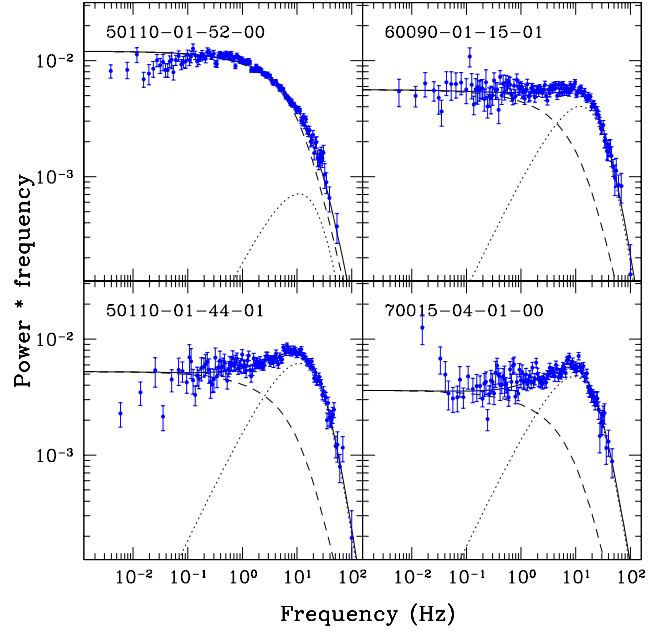}}
\end{center}
\caption{The observed power spectra and best fit models (solid curves; fitted at $f>0.1$ Hz) for four representative data sets. The dotted and dashed curves give the linear and non-linear components, respectively.  }
\label{f:fits}
\end{figure}

\begin{figure}
\begin{center}
{\includegraphics[width=0.9\linewidth,angle=0]{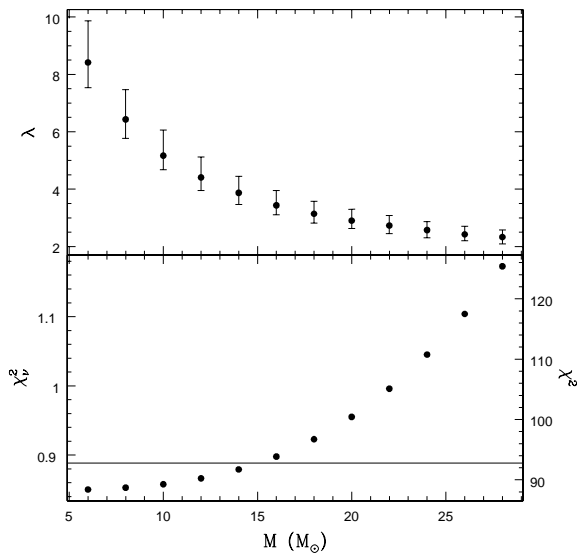}}
\end{center}
\caption{ The variations of $\chi^2_\nu$  and the best fit damping factor, $\lambda$, vs.\ the black-hole mass for the observation 50110-01-52-00 (see Fig.\ \ref{f:fits}), which is one of the two data sets yielding the lowest $M_{\rm u}$.   At $M > M_{\rm u}\simeq 16 \msun$, $\Delta \chi^2 > 2.71$ with respect to the overall $\chi^2$ minimum, as shown by the solid horizontal line in the bottom panel. Hence, $16 \msun$ is an upper limit for $M$. The corresponding constraint for the damping factor is $\lambda \ga 3$. }
\label{f:chi2}
\end{figure}

The spectra were then fitted to the damped harmonic oscillator model for $f\geq 0.1$ Hz. The resulting best fit parameters and their 90 per cent uncertainties (i.e., for $\Delta \chi^2 = +2.71$) are given in Table \ref{t:fits}. Four representative spectra and their fits are shown in Fig.\ \ref{f:fits}. As mentioned in Section \ref{model}, there is a near degeneracy between the black hole mass and the damping factor, $\lambda$, in all cases. Therefore, the fits in Fig.\ \ref{f:fits} are for $M = 10 \msun$. Generally, the best fits are obtained for low masses, and we can obtain only upper limits, $M_{\rm u}$, using the criterion of $\Delta \chi^2 = +2.71$ with respect the overall best fit for each data set. As we show in Table \ref{t:fits}, we find $M_{\rm u}$ in the range of (16--36)$\msun$, though $M_{\rm u}\la 20\msun$ in most cases. We illustrate this degeneracy in Fig.\ \ref{f:chi2}, where we plot the variations of the reduced $\chi^2$ and of $\lambda$ vs.\ $M$ for one of the two data sets yielding the lowest $M_{\rm u}\approx 16\msun$.  Thus, the limit of $M \la 16 \msun$ is satisfied by all of the data, and we can consider it as our resulting overall limit. On the other hand, $M \la 20 \msun$ is satisfied by most of our data. 

The mass of the black hole in Cyg X-1 determined using various other techniques remains relatively uncertain. \citet{gies86} found $M\simeq 16\pm 5\msun$, consistent with the later determination by \citet{Zio05} of $13\la M/\msun \la 29$. On the other hand, \citet{herrero95} , based on modeling of the atmosphere of the companion star, obtained $M\simeq 10\pm 5\msun$. Still, those determinations are consistent with our limit of $M\la (16$--$20) \msun$.

As seen in Fig.\ \ref{f:chi2}, the fit in the shown case yields the damping factor of $\lambda \ga 3$. On the other hand, we can use the independent limits on $M$ to constrain $\lambda$ from above. In any case, $M>5\msun$ \citep{herrero95}, or, more likely, $\ga 6.5\msun$ \citep{Gie99}. The latter implies $\lambda \la 8$. These constraints are consistent with Table \ref{t:fits}, which shows $\lambda\approx 5\pm 1$ for $M=10\msun$, taking into account that our fits constrain mostly $\lambda/M$ (Section \ref{model}).  

\begin{table}
\centering
  \caption{Results of the fits to soft-state power spectra of Cyg X-1. The values and errors of $\lambda$, are given for $M = 10 \msun$; those for fractional rms and $q_{\rm nl}$ are approximately independent of $M$.  }
\begin{tabular}{cccccc}
\hline
Observation ID & $\lambda$ & rms (\%) & $q_{\rm nl}$ & $\chi^2_\nu$ & $M_{\rm u}/\msun$ \\
\hline
10512-01-09-00& $ 5.22^{+ 0.38}_{- 0.35}$ & $ 24^{+1}_{- 1}$ & $ 0.46^{+ 0.04}_{- 0.03}$ & 1.27 & $20$   \\
10512-01-09-01& $ 5.02^{+ 0.20}_{- 0.20}$ & $ 32^{+1}_{-1}$ & $ 0.41^{+ 0.03}_{- 0.02}$ & 1.10 & $18$   \\
50110-01-52-00& $ 5.17^{+ 0.89}_{- 0.49}$ & $ 36^{+1}_{-1}$ & $ 0.13^{+ 0.04}_{- 0.02}$ & 0.86 & $16$   \\
50110-01-44-01& $ 5.12^{+ 0.20}_{- 0.25}$ & $ 27^{+1}_{-1}$ & $ 0.57^{+ 0.04}_{- 0.03}$ & 0.89 & $18$   \\
50110-01-53-00& $ 4.36^{+ 1.24}_{- 0.71}$ & $ 28^{+1}_{-1}$ & $ 0.15^{+ 0.06}_{- 0.03}$ & 1.13 &$18$   \\
60089-03-01-01& $ 4.45^{+ 0.82}_{- 0.80}$ & $ 35^{+1}_{-1}$ & $ 0.22^{+ 0.04}_{- 0.04}$ & 0.82 & $24$   \\
60090-01-07-01& $ 6.31^{+ 0.45}_{- 0.49}$ & $ 24^{+1}_{-1}$ & $ 0.48^{+ 0.04}_{- 0.04}$ & 1.30 & $28$   \\
60090-01-11-00& $ 5.43^{+ 0.69}_{- 0.51}$ & $ 36^{+1}_{-1}$ & $ 0.13^{+ 0.03}_{- 0.02}$ & 1.16 & $16$   \\
60090-01-12-00& $ 4.97^{+ 1.15}_{- 0.65}$ & $ 33^{+1}_{-1}$ & $ 0.15^{+ 0.06}_{- 0.03}$ & 0.84 & $18$   \\
60090-01-15-01& $ 4.63^{+ 0.24}_{- 0.27}$ & $ 27^{+1}_{-1}$ & $ 0.44^{+ 0.03}_{- 0.03}$ & 0.77 & $22$   \\
70015-04-01-00& $ 5.94^{+ 0.43}_{- 0.40}$ & $ 23^{+1}_{-1}$ & $ 0.61^{+ 0.06}_{- 0.05}$ & 1.27 & $36$   \\
\hline
\end{tabular}
\label{t:fits}
\end{table}

\section{The rms-flux relationship}
\label{rms-flux}

Above, we have shown that our model fits well the soft-state power spectra. Another major observable describing variability of accreting sources is the flux rms as a function of the flux, which has been shown to be linear in accreting black-hole and neutron-star systems \citep{uh01} and AGNs (e.g., \citealt{m07}). In particular, the fractional rms $\simeq$ constant in both the hard and soft states of Cyg X-1 \citep{gleissner04}. 

We point out here that, in our formalism, this linearity is an intrinsic property of the dynamic quantity governing the variability, $x(t)$, but it is not related to the relationship between the flux and $x(t)$. The flux variance over a time interval, $T$, is 
\begin{equation}
\sigma^2={1\over T}\int_0^T \left[S(t)-S_0\right]^2{\rm d}t,
\label{sigma}
\end{equation}
or the corresponding sum over some time steps, $\Delta t$. If the normalized flux is $s(t)=a\left[x^n(t)-\langle x^n(t)\rangle\right]$, the flux variance becomes,
\begin{equation}
\sigma^2={S_0^2\over T}\int_0^T \left[x^n(t)-\langle x^n(t)\rangle \right]^2{\rm d}t,
\label{sigma_n}
\end{equation}
where $n=1$ in the linear case, and $n\geq 2$ in the non-linear one. Then, if the variance of $x(t)$ [given by the integral in equation (\ref{sigma_n}) for $n=1$] is approximately the same in different time intervals, $T$, i.e., $\sigma(X)\propto X_0$ in each time interval, it is easy to see that $\sigma(S)\propto S_0$ at any $n$, with only the proportionality constant changing. As an illustrative example, let's take a single harmonic of $x(t)$, i.e., $x(t)=\sin f t$, for which $\sigma^2=S_0^2/2$ and $S_0^2/8$ for $n=1$ and 2, respectively, i.e., the rms-flux relationship remains linear even when the flux response to the underlying variability is non-linear.

In our model, the power is supplied to the disc via the driving term, $C(t)$. As $x(t)$ is linearly related to $C(t)$, the observed rms-flux linearity can occur if the standard deviation of $C(t)$ is proportional to its local average. Since we do not specify the nature of $C(t)$ [except for assuming $\tilde C(f) \propto 1/f$], our model does not imply a particular rms-flux relation, but it is consistent with it being linear. 

\section{Discussion and conclusions}
\label{discussion}

We have developed a relatively simple accretion disc variability model, in 
which we assumed that the local dynamic variability at each radius is governed by a damped harmonic oscillator equation at the corresponding epicyclic frequency. This variability induces both linear and non-linear (quadratic in our model) responses in the emitted photons. The observed variability is then the sum of those from different radii weighted by the rate of local energy dissipation. The power spectrum predicted by our model is characterized
by four parameters, which are the total fractional rms, the ratio of the non-linear power to the linear one, $q_{\rm nl}$, the mass of the black hole, $M$, and the damping factor, $\lambda$. We find that $M$ and $\lambda$ are closely correlated, and we can determine only $\lambda/M$. Thus, the predicted spectrum is a function of only three effective parameters. We find we can well explain the different types of the power spectra observed from Cyg X-1 in the soft state with this model. Although the degeneracy in $\lambda$ and $M$ does not allow an unambiguous determination of $M$, we are able to obtain an upper limit of $M \la (16$--$20) \msun$. Remarkably, this limit is close to the range of other mass measurements of Cyg X-1. The corresponding constraint on the damping factor is $3\la \lambda\la 8$. 

This model should, in the future, be made consistent with a radiative model for the variability. Whereas we here assume that the dynamic disc variability
induces linear and non-linear responses in the emitted radiation, a
radiative model should identify the parameters of the emitting medium (e.g., the optical depth, the plasma temperature, etc.) that drive the variability. It should predict the rms vs.\ the photon energy and time lags between different energies. One important issue we need to understand within such framework is the nature of the non-linear response of the plasma, which inherent presence in the soft state of Cyg X-1 we have found from the observations. On the theoretical side, we can state that its relative strength, $q_{\rm nl}$, depends on the frequency range, $\Delta f_{\rm coh}$, over which the stochastic driver $C(t)$ is coherent. If the nature of the stochastic driver does not change, variations in the strength of the non-linear component imply that the non-linear response of the plasma varies. This predicts $q_{\rm nl}$ to be correlated with the parameters of the hot plasma determining its emission. However, quantitative predictions can be made only when a comprehensive radiative model for the variability is available.

An important constraint on $C(t)$ is that it should preserve the observed linear relationship between the rms and the flux. As we have shown, this implies that the variance of $C(t)$ remains approximately constant over time. On the other hand, we have shown that even a non-linear response of the flux to the underlying dynamic variability preserves the linearity of the rms-flux relation.

\section*{Acknowledgments}

We are grateful to Marek Gierli\'nski for providing us with his viewgraphs and an animation illustrating variability of Cyg X-1, and to the referee for a stimulating comment. This research has been supported in part by the Polish MNiSW grant NN203065933 (2007--2010), the Polish Astroparticle Network 621/E-78/SN-0068/2007, and the exchange program between the Polish Academy of Sciences and the Indian National Science Academy.

\label{lastpage}

\end{document}